	\documentclass[prd,aps,showpacs,tightenlines,nofootinbib,preprint]{revtex4}

\usepackage[dvips]{graphicx}
\usepackage{amsmath,amssymb}
\usepackage{color}

\def\mr{\mathrm}

\def\nm{\nonumber}

\def\pd{\partial}

\begin{document}

\title{Combined features in the primordial spectra induced by a sudden turn in two-field DBI inflation}
\author{Shuntaro Mizuno$^{1}$, Ryo Saito$^{2}$, and David Langlois$^{2}$}
\affiliation{$^1$Waseda Institute for Advanced Study, Waseda University, 
	Tokyo 169-8050,Japan,\\
$^2$APC, (CNRS-Universit\'{e} Paris 7),
        10 rue Alice Domon et L\'{e}onie Duquet, 75205 Paris, France\\
}

\begin{abstract}
We investigate the features generated by  a sharp turn along the inflationary trajectory in a two-field  model of Dirac-Born-Infeld inflation, where one of the fields  is heavy. Distinct features are generated by two different effects: the mixing of the light and heavy modes during the turn, on the one hand, and   the resonance between the oscillations along the heavy direction after the turn, on the other hand. Contrary to  models with standard kinetic terms, the resonance effect is not strongly suppressed because the action contains derivative interactions. Working in the potential basis,  we study the oscillations after the turn and compute the amplitude  of the mixing and resonance features in the power spectrum, as well as in the bispectrum for the latter effect. We find that the amplitudes and positions of these combined features obey specific consistency relations, which could be confronted with cosmological data.

\end{abstract}

\maketitle

\section{Introduction}\label{sec:introduction}
 The detailed analysis of primordial density fluctuations represents the main window onto 
high energy physics at work during  cosmic inflation. 
In the inflationary scenario, the primordial fluctuations  are generated from the quantum fluctuations of 
one (or several) scalar field(s), called the inflaton(s), 
characterized by  a light mass, i.e.  smaller than the Hubble parameter,  during  inflation 
(see e.g. \cite{Langlois:2010xc} for a pedagogical introduction). 
Primordial fluctuations are observable today via their imprint on  the cosmic microwave background (CMB) and large scale structures. 
 
 From the perspective of inflationary model building,  light scalar fields are usually not the only scalar fields present. In general, one also finds  scalar fields with masses of the order of the Hubble parameter, or even much heavier.  
 However, in contrast  with inflatons (i.e. light modes),  these heavy fields (heavy modes) are usually considered as irrelevant from an observational point of view as  their fluctuations are significantly suppressed 
on cosmological scales.
 
 Recently, however, it was pointed out that such heavy modes can in fact affect 
the fluctuations generated during inflation when the inflationary trajectory in field space is bent. Indeed, in such situations, the heavy field can be displaced from the minimum of the potential  as a consequence of  the centrifugal force induced  by the turn of the trajectory. For a moderate bending of the trajectory, it was shown that
the system is described  as an effective single light field model with 
a reduced speed of sound \cite{Tolley:2009fg,Achucarro:2010jv,Achucarro:2010da}
(see also \cite{Gao:2009qy,Chen:2009we,Chen:2009zp,Cremonini:2010ua,
Peterson:2011yt,Behbahani:2011it,Achucarro:2012sm,Avgoustidis:2012yc,Chen:2012ge,Pi:2012gf,Achucarro:2012yr,Burgess:2012dz,Gwyn:2012mw,Noumi:2012vr,Battefeld:2013xka,Cespedes:2013rda,Gao:2013zga,Emami:2013lma,Castillo:2013sfa}  for other works investigating the effects of heavy modes in terms of an effective theory for the light field).
Moreover, for a sufficiently sharp bending of the trajectory, 
the heavy field oscillates around its minimum after the turn and
the effective single light field description is no longer valid.

A sharp turn can induce two types of features, which are potentially observable in the primordial spectra such as the power spectrum and bispectrum. The first type of features is induced around the scale that crosses the Hubble scale at the time of the turn through a large mixing between the light and heavy modes during the turn \cite{Shiu:2011qw,Cespedes:2012hu,GLM,Konieczka:2014zja}.
In addition to this, when the turn is sufficiently sharp to induce oscillations of the heavy field, another type of features is generated. The latter appear around the scale that crosses the mass scale
of the heavy field at the time of the turn through the resonance between the oscillations in the background trajectory and the inflaton fluctuations 
\cite{Chen:2008wn,Chen:2011zf,Chen:2011tu,Saito:2012pd}.
Although it was shown in Refs~\cite{GLM2,Noumi:2013cfa} 
that this resonance effect is not efficient for canonical scalar fields, 
the features can nevertheless be large when the light and heavy modes are coupled via derivative interactions \cite{Saito:2012pd} (see also \cite{Kobayashi:2012kc}). 

Detection of a combination of both types of features  could provide compelling  evidence for
the existence of such heavy modes during inflation. Indeed, with only a single type of features analyzed, 
it is difficult to obtain a large statistical significance  because one cannot predict the position of the feature, which depends on when the turn occurred during inflation.
However, it  is possible to increase the significance of the signal  if different and correlated  features can be detected in the power spectrum and bispectrum 
\cite{Achucarro:2012fd,Achucarro:2013cva,Gong:2014spa,Achucarro:2014msa}.
This is the case here, as
   the relative position between mixing and resonance  features  is determined by the ratio between the heavy mass and the Hubble parameter.

The purpose of this paper is to investigate how much the heavy field can be excited during a  sharp turn in a model with derivative interactions, so that both mixing and resonance features  are expected. A model of inflation that naturally provides such type of interactions is based on the Dirac-Born-Infeld (DBI) Lagrangian 
with multiple fields \cite{Langlois:2008wt,arXiv:0806.0336,Arroja:2008yy}. 
Whereas derivative interactions  were treated perturbatively in the previous work \cite{Saito:2012pd}, DBI inflation automatically includes  higher-order terms,  which could become important at the turn. 
In this framework, the goal of this work is to  determine the efficiency of the heavy mode excitation, depending on the model parameters,  and to establish a  relation between the features due to the mixing and resonance effects.

The organization of this paper is as follows. In Section~\ref{sec:BG}, we introduce our two-field model and derive the background equations of motion, which are reexpressed in a convenient basis. In Section \ref{sec:EX}, we concentrate on the evolution of the trajectory just after the turn, in particular on its oscillations due to the excitation of the heavy mode. After deriving analytical results, we study numerically a specific example. Section \ref{sec:Features} is devoted to the features generated in the power spectrum by the mixing effect and the resonance effect. Features in the bispectrum are also discussed. We summarize our results in the final section. In the appendix, we give some details about  the resonance feature in the power spectrum.

\section{Background evolutions}\label{sec:BG}
In this section, we introduce the  two-field model of DBI inflation that  we will study, and derive the relevant background equations of motion.

\subsection{Two-field DBI inflation}

Following the previous works \cite{Langlois:2008wt,arXiv:0806.0336,Arroja:2008yy}, 
we consider a model with 
two scalar fields $\phi^I$ ($I=1,2$), governed by the action
    \begin{equation}
    \label{eq:action}
        S=\int d^{4}x\sqrt{-g}\,  P(X^{IJ},\phi^{I})\,, 
    \end{equation}
where $P$ is a function of the scalar fields and of their kinetic terms
\begin{equation}
 X^{IJ} \equiv - \frac{1}{2} \partial _{\mu} \phi^{I} \partial ^{\mu} \phi^{J}\,,
\end{equation}
and $g$ is the determinant of the spacetime metric $g_{\mu\nu}$.

For  two-field DBI inflation models, $P(X^{IJ},\phi^{I})$ is explicitly given by
\begin{equation}
 P(X^{IJ},\phi^{I}) = \tilde{P}(\tilde{X},\phi^{I}) = - \frac{1}{f(\phi^{I})} \left(\sqrt{1-2 f(\phi^{I}) \tilde{X}} - 1\right) - V\left(\phi^{I}\right),
\label{multifieldaction}
\end{equation}
where $f(\phi^I)$ and $V(\phi^I)$ are functions of the scalar fields.
Here, $\tilde{X}$ is defined in terms of the determinant (we use  Einstein's implicit summation rule for the scalar field indices)
\begin{eqnarray}
\mathcal{D} &=& \mbox{det} (\delta^{I}_{J} - 2 f X^{I}_{J} )\nonumber \\
&=& 1 - 2 f G_{IJ} X^{IJ} + 4 f^{2} X^{[I}_{I} X^{J]}_{J}\,,
\label{eq:determinant}
\end{eqnarray}
as
\begin{equation}
 \tilde{X} = \frac{(1-\mathcal{D})}{2 f},
\end{equation}
where
$G_{IJ}$ is the metric in the field space. In the context of string theory, the DBI action describes the effective dynamics of a D3 brane in a higher-dimensional background spacetime and  $f(\phi^{I})$ is related to 
 the warp factor  $h(\phi^{I})$ of this higher-dimensional spacetime and the brane tension $T_3$ as
\begin{equation}
 f(\phi^{I}) \equiv \frac{h(\phi^{I})}{T_{3}}\,.
\end{equation}

Note that the action (\ref{eq:action}) contains derivative interactions such as $(X^{12})^2$ and $X^{11}X^{22}$, which were considered in Ref.~\cite{Saito:2012pd}, as well as self couplings and higher-order terms.
 In the DBI inflation model, the magnitude of all these couplings is determined by the single function $f(\phi^I)$, or equivalently by the effective sound speed $c_s$ defined below. A small  speed of sound corresponds to large derivative interactions.
 
\subsection{Background equations of motion}

In a spatially homogeneous and isotropic spacetime, endowed  with the  metric
\begin{equation}
ds^2=g_{\mu\nu}dx^\mu dx^\nu=-dt^2+a^2(t)\delta_{ij}dx^idx^j,
\end{equation}
the evolution of  the scale factor $a(t)$ is governed by  the Friedmann equations
   \begin{equation}
H^{2}=\frac{1}{3}\left(\frac{1}{(1+c_s) c_s} G_{IJ}\dot{\phi}^{I}\dot{\phi}^{J}+V\right),\qquad\qquad \dot{H}=-\frac{1}{2 c_s}G_{IJ}\dot{\phi}^{I}\dot{\phi}^{J}\equiv-H^{2}\epsilon,\label{eom_bg}
\end{equation}
where $H\equiv \dot a/a$ is the Hubble parameter and a dot denotes a derivative with respect to the cosmic time $t$. We use  units such that $M_P \equiv (8\pi G)^{-1/2}=1$. 
Here, $c_s$ is the effective sound speed, corresponding to the propagation speed of the perturbations, and is given by \cite{Langlois:2008wt}
\begin{equation}\label{eq:c_s}
 c_{s} \equiv \sqrt{\frac{\tilde{P}_{,\tilde{X}}}{\tilde{P}_{,\tilde{X}} + 2 \tilde{X} \tilde{P}_{,\tilde{X} \tilde{X}}}} = \sqrt{1- 2 f \tilde{X}},
\end{equation}
where $,_{\tilde{X}}$ means the partial derivative with respect to $\tilde{X}$. Note that $\tilde{X}$ coincides with $X \equiv G_{IJ} X^{IJ}$ in the homogeneous background because all spatial derivatives vanish. From the action (\ref{multifieldaction}),
we can show that
\begin{equation}
 \tilde{P}_{,\tilde{X}} = \frac{1}{c_{s}}.
\end{equation}

By introducing the components of the acceleration in curved space
(here  the field space), which we can write as
\begin{equation}
 \mathcal{D}_t \dot{\phi}^{I} \equiv \ddot{\phi}^I + 
\Gamma^I _{JK} \dot{\phi}^J \dot{\phi}^K\,,
\end{equation}
and the simplified notation ${}_{,J}\equiv \partial{}/\partial \phi^J$, 
the equations of motion for the homogeneous scalar fields are
\begin{eqnarray}
&&\mathcal{D}_t \dot{\phi}^{I}
+(3-\epsilon_s)H\dot{\phi}^{I}
-c_s G^{IJ} \tilde{P}_{,J}=0\label{phi_eom_pb}\,;\quad
\tilde{P}_{,J} 
= -V_{,J} +\frac{(1-c_s)^2}{2 c_s} \frac{f_{,J}}{f^2}\,,\;\;\;\;
\epsilon_s \equiv \frac{\dot{c}_s}{H c_s},
\label{eqs_fields_gen}
\end{eqnarray}
or, in an even more compact form,
\begin{eqnarray}
a^{-3} \mathcal{D}_t \left( a^3 \frac{1}{c_s} \dot{\phi}_I \right)
= \tilde{P}_{,I}\,, 
\end{eqnarray}
where we have used the field space metric $G_{IJ}$ to lower the field index $I$,
so that $\dot{\phi}_I \equiv G_{IJ} \dot{\phi}^J$.
$ \mathcal{D}_t$ acts as an ordinary time derivative on field space scalars
(i.e. quantities without field space indices) and $\mathcal{D}_t G_{IJ} =0$.

Since our purpose is  to focus on the effects of a reduced sound speed
during the turn and on the resonance associated with the background oscillation after it, from now on, we consider a  simplified  model characterized by $G_{IJ}=\delta_{IJ}$ and $f=\text{const}$.
This implies  $\tilde{P}_{,I}=-V_{,I}$, 
which simplifies Eq.~(\ref{eqs_fields_gen}) as
	\begin{equation}
		\ddot{\phi}^I +(3-\epsilon_s) H \dot{\phi}^I + c_s V^{,I}=0\,. \label{eqs_fields_simp}
	\end{equation}
From this equation, one sees that the sound speed, which is always smaller than $1$, has two consequences: a modification of the Hubble friction and a flattening of the effective potential felt by the scalar fields.

\subsection{Kinematic and potential bases}
We will now solve 
the evolution equation (\ref{eqs_fields_simp}) following Ref. \cite{GLM}, including the new effects due to the nontrivial sound speed. First of all, let us define the light and heavy directions in field space, corresponding to  the basis $e^I_m$ that diagonalizes the Hessian matrix of the potential, i.e. such that
	 \begin{equation}{\label{V_mn_diag}}
        V_{,mn} \equiv e^I_m e^J_n\,  V_{,IJ}= \mathrm{diag}\{m_{l}^2, m_{h}^2\}\,.
    \end{equation}
Because of the flattening of the potential due to the sound speed, the masses are effectively reduced by a factor $\sqrt{c_s}$.
We use the terms ``light" and ``heavy" for these effective masses and assume the hierarchy 
	\begin{align}\label{eq:mass_hierarchy}
		\sqrt{c_s}m_l \ll H \ll \sqrt{c_s}m_h\,.
	\end{align}
Following the terminology introduced in Ref. \cite{GLM}, we will call the basis that diagonalizes the Hessian matrix of the potential, the ``potential basis".
In a similar fashion, we call the basis 
associated with
 the usual adiabatic-entropic decomposition, the ``kinematic basis".

To describe the deviation between the velocity (adiabatic) direction
	\begin{equation}
		n^I= \frac{\dot\phi^I}{\dot\sigma}, \qquad \dot\sigma\equiv \sqrt{\delta_{IJ}\, \dot\phi^I\dot\phi^J},
	\end{equation}
and the light direction, we consider the evolution of the angle $\psi$ between  the adiabatic and  light directions.
 The components of the adiabatic unit vector in the potential basis  can be explicitly expressed in terms of $\psi$ as
 	 \begin{equation}{\label{psi_def}}
	 	\{n_1,n_2\} = \{\cos\psi,\sin\psi\}\,.
	\end{equation}
The angle $\psi$ can be written as the difference
	\begin{eqnarray}
		 \psi = \theta_k - \theta_p\,,
	 \label{rel_psi_theta_p}
 	\end{eqnarray}
where $\theta_k$ and $\theta_p$ are, respectively, the angles of the kinematic and potential bases with respect to the original field basis.
While the angle $\theta_p$ can be determined by the shape of the potential, we should solve Eq. (\ref{eqs_fields_simp}) to know the evolution of the other angle $\theta_k$. 
Moving to the kinematic basis, the equations of motion (\ref{eqs_fields_simp}) can be rewritten as
    \begin{align}
        (3+\epsilon_\sigma -\epsilon_s) H\dot{\sigma} + c_s V_{,\sigma} &= 0\,, \qquad V_{,\sigma} \equiv n^I V_{,I}\,,  \label{bckgd_adiab} \\
	 \dot\sigma\dot\theta_k + c_s V_{,s} &=0\,, \qquad V_{,s} \equiv s^I V_{,I}\,,
\qquad \epsilon_\sigma \equiv \frac{\ddot{\sigma}}{H \dot{\sigma}}\,.
\label{bckgd_enab}
    \end{align}
Combining these equations yields the evolution equation for $\theta_k$,
	\begin{equation}{\label{theta_2nd_eom_gen}}
        		\ddot{\theta}_k+(3+2\epsilon_\sigma-2\epsilon_s) H\dot{\theta}_k
+c_s V_{,\sigma s}=0\,.
    	\end{equation}
Hence, substituting the expression (\ref{rel_psi_theta_p}) into Eq. (\ref{theta_2nd_eom_gen}), we obtain
	  \begin{equation}{\label{theta_2nd_eom_2f}}
        \ddot{\psi}+ 3H\left(1+\frac23 
(\epsilon_{\sigma}-\epsilon_s )\right)\dot{\psi}+ \frac{c_s}{2} \left(m_h^2-m_l^2 \right)\sin(2\psi)= - \ddot{\theta}_p - 3H\left(1+\frac23 
(\epsilon_{\sigma}-\epsilon_s )\right)\dot{\theta}_p\,.
    \end{equation}
    In order to obtain simple analytical solutions, we will neglect $\epsilon_\sigma-\epsilon_s$ in Eq. (\ref{theta_2nd_eom_2f}). 
    Because the velocity contains a contribution from the heavy field $\dot{\phi}_h$, its derivative induces a large factor $m_h$.
    Evaluating these contributions, we find
    	\begin{align}
		\epsilon_{\sigma} - \epsilon_{s} \sim \frac{\sqrt{c_s}m_h}{H}\left(\frac{\dot{\phi}_h}{\dot{\sigma}}\right)^2,
	\end{align}
using $(\dot{\phi}_h/c_s)^{\cdot} \sim m_h^2 \phi_h$ and $\dot{\phi}_h \sim \sqrt{c_s}m_h\phi_h$, 
which can be shown to be consistent with the evolution equation (\ref{eqs_fields_simp}) for the heavy scalar field with $\sqrt{c_s}m_h \gg H$. 
As will be shown in Sec. \ref{sec:Features}, 
this is the same order as the amplitude of the features in the power spectrum induced by the sharp turn.
Therefore, $\epsilon_\sigma-\epsilon_s$ can be safely  neglected  for a reasonable modulation of the power spectrum.
We also assume that the angle $\psi$ is sufficiently small so that the sine can be replaced by its argument.
    With these approximations and by neglecting $ m_l^2$, the evolution equation (\ref{theta_2nd_eom_2f}) reduces to
    \begin{equation}{\label{psi_eom_sim}}
        \ddot{\psi}+ 3H\dot{\psi}+ c_s m_h^2\psi = 
- \ddot{{\theta}}_p - 3H\dot{{\theta}}_p\,.   
    \end{equation}

    It is worth mentioning that when the condition for inflation
    	\begin{align}
		\epsilon \equiv -\frac{\dot{H}}{H^2} = \frac{\dot{\sigma}^2}{c_s H^2} \ll 1,
	\end{align}
    as well as $\epsilon_{\sigma}-\epsilon_s \ll 1$ are satisfied,
    the sound speed $c_s$ is approximately determined by the potential as
    \begin{eqnarray}
    	c_{s} \simeq \sqrt{\frac{1}{1 + 2\epsilon_V f H^2}} \equiv c_{s0}\,, \qquad \epsilon_V \equiv \frac{1}{2}\left(\frac{V_{,\sigma}}{V}\right)^2\,,
    \label{cs_approximation}
    \end{eqnarray}
and $\epsilon \simeq c_s \epsilon_V$.

\section{Excitation of the heavy mode}
\label{sec:EX}

  \subsection{Efficiency of the excitation}   
   Assuming that $H$ and $\sqrt{c_s}m_h$ remain approximately constant during the turn, 
 we can formally solve Eq. (\ref{psi_eom_sim}) as
\begin{equation}{\label{psi_sol_gf}}
        \psi (t)= -\int^t dt'\, G(t,t') \left[ \ddot{{\theta}}_p(t') + 3H\dot{{\theta}}_p(t') \right],
    \end{equation}
using the retarded Green's function given by
    \begin{equation}{\label{psi_green_fun}}
        G(t,t') = \Theta(t-t') \frac{\sin (\omega (t-t'))}{\omega} e^{-\frac{3}{2}H(t-t')}   ,\qquad\qquad \omega = \sqrt{c_s m_h^2- \frac{9}{4}H^2},
    \end{equation}
    where $\Theta$ is the Heaviside distribution.
Because of 
the assumption (\ref{eq:mass_hierarchy}), the frequency $\omega$ can be approximated by
	\begin{align}\label{eq:omega}
		\omega \simeq \sqrt{c_s}\, m_h\,.
	\end{align}
  
  The precise evolution of the angle $\theta_p(t)$  depends on the details of the potential. In order to work with analytical expressions, here, we simply characterize the turn by two parameters, as in  \cite{GLM}, 
  and use the function
  	\begin{equation}
    		\label{thetapdot}
        		\dot{{\theta}}_p(t) = \Delta\theta\frac{\mu }{\sqrt{2\pi}} e^{-\frac{1}{2}\mu^2t^2}\,,
	\end{equation}
where the turn is assumed to occur at time $t=0$. 
$\Delta\theta$ represents the global variation of the angle during the turn and $\mu$ is related to the duration of the turn by $\Delta t_{\rm turn}\sim \mu^{-1}$.
When the slow-roll conditions are satisfied, $\mu$ can be estimated as
	\begin{align}\label{eq:mu_pot}
		\mu \sim s \dot{\sigma} \sim s c_s \sqrt{\epsilon_V} H,
	\end{align}
where $s$ represents the ``sharpness'' of the turn in  field space: $s \sim (\partial \theta_p/\partial \sigma)/\Delta \theta$.

Plugging the expression (\ref{thetapdot})  into Eq. (\ref{psi_sol_gf}), we get
    \begin{eqnarray}
\psi(t)
 & = & -\frac{\Delta\theta}{2} \sqrt{1+\frac{9H^2}{4\omega^2}} e^{-\frac{3}{2}Ht}\, \Re\left[e^{i\alpha+\varphi^{2}/2}e^{-i\omega t}\mathrm{erfc}\left(-\frac{\mu t-\varphi}{\sqrt{2}}\right)\right],\label{psi_ana}
\end{eqnarray}
where $\Re$ denotes the real part of the argument, $\mathrm{erfc}(z) \equiv 1- \mathrm{erf}(z)$ is the  complementary error function and we have  introduced the parameters
    \begin{equation}{\label{alpha_varphi_def}}
        \alpha := \arctan\left( \frac{3H}{2\omega} \right),\qquad\varphi := \frac{\omega}{\mu}\sqrt{1+\frac{9H^{2}}{4\omega^{2}}}\, e^{i\left(\frac{\pi}{2}-\alpha\right)}\,.
    \end{equation}
    
In the following,
we will be  mainly interested in  sharp turns, corresponding to  $\mu/\omega \gtrsim 1$, which lead to post-turn  oscillations.
In this case, the expression (\ref{psi_ana}) is   approximated by
            \begin{equation}{\label{psi_sharp_app}}
                \psi(t) \approx -\frac{\Delta\theta}{2}e^{-\frac{\omega^{2}}{2\mu^{2}}}\mathrm{erfc}\left(-\frac{\mu t}{\sqrt{2}}\right)e^{-\frac{3}{2}Ht}\cos\left(\omega t-\alpha-\frac{3H\omega}{2\mu^{2}}\right).
            \end{equation}
After the turn, i.e. for $\mu t \gtrsim 1$,  $\psi$ starts to oscillate around the light direction,  with the damping factor $e^{-\frac{3}{2}H t}$. In terms of $\psi$, the evolution of  the heavy field is given by 
	\begin{align}
		\dot{\phi}_h = \dot{\sigma}\sin\psi \simeq  \dot{\sigma}\psi\,,
	\end{align}
	which expresses the oscillations of the heavy scalar field excited by the turn. 
	
Since it is reasonable to consider that the energy of the excited oscillations is mainly extracted  from the kinetic energy before the turn, 
we define the efficiency of the excitation through 
	\begin{equation}\label{E_r_def}
                \xi_{\rm osc} \equiv \left( \frac{\dot{\phi}_{h,{\rm max}}}{\dot{\sigma}} \right)^2 \simeq (\psi_{\rm max})^2,
            \end{equation}
where we have assumed  in the second equality that the velocity $\dot{\sigma}$ is approximately constant during the turn (which will be verified in the explicit example we consider later).
Using the expression (\ref{psi_sharp_app}), we  find that $\xi_{\rm osc}$ is given by
            \begin{equation}\label{E_r_est}
                \xi_{\rm osc} \simeq e^{-\frac{\omega^{2}}{\mu^{2}}} (\Delta \theta)^2 \,.
            \end{equation}
In general, the prefactor $e^{-\omega^2/\mu^2}$ changes if we use a  time dependence for $\dot{\theta}_p$  that differs from  Eq. (\ref{thetapdot}). 
However, in the limit $\mu\gg\omega$, the efficiency reaches  its maximum value
	\begin{align}\label{eq:eff_max}
		\xi_{\rm osc,max} \simeq (\Delta \theta)^2 \,,
	\end{align} 
which only depends on the global variation of the angle, $\Delta \theta$.

Using the expressions (\ref{eq:omega}) and (\ref{eq:mu_pot}), the condition $\mu \gg \omega$ can be expressed as
	\begin{align}\label{eq:sharp_turn}
		\frac{m_h}{H} \ll s\sqrt{c_s\epsilon_V} \simeq s\sqrt{\epsilon}\,.
	\end{align}
Thus, in terms of the slow-roll parameter $\epsilon$, the sound speed does not appear in the above condition. 
However, for a given potential, a small sound speed makes the heavy mode more difficult to  excite.
This is because, though the heavy mass is suppressed by the small sound speed, the incident velocity, $\dot{\sigma}$, is reduced as well.
By contrast,
 the maximum efficiency (\ref{eq:eff_max}) does not depend on the sound speed but only on the global variation of the angle $\Delta \theta$.

For completeness, let us also briefly mention the analytic result in the soft turn case, even if it does not induce any resonance.
In the limit $\mu\ll \omega$, the evolution of $\psi$, before and during the turn,  is approximately given by
        \begin{equation}{\label{psi_soft_app}}
            \psi(t) \approx \frac{\Delta\theta}{\sqrt{2\pi}}\,\frac{\mu^2}{\omega^{2}}\, e^{-\frac{1}{2}\mu^{2}t^{2}}\left( \mu t-3\frac{H}{\mu}\right)\,.
        \end{equation}        
Moreover,  long \emph{after} the turn, when  $\mu t \gtrsim \omega/\mu\gg 1$, $\psi$ behaves like
            \begin{equation}
                \psi\left(t\right)\approx-\Delta\theta\left[e^{-\frac{\omega^{2}}{2\mu^{2}}}e^{-\frac{3}{2}Ht}\cos\left(\omega t-\alpha-\frac{3H\omega}{2\mu^{2}}\right)-\frac{1}{\sqrt{2\pi}}e^{-\frac{1}{2}\mu^{2}t^{2}}\frac{1}{\mu t}\right].
            \end{equation}
This shows that, after the turn, the heavy mode oscillates around the minimum,  but with a tiny amplitude since $\omega/\mu\gg 1$.

\subsection{Numerical analysis}

In the previous subsection, we have estimated analytically the efficiency of the excitation by assuming $\sqrt{c_s} \, m_h$ to be constant during the turn. In this subsection, we consider an explicit model and compute numerically the efficiency parameter $\xi_{\rm osc}$, in order to compare the result with our analytical estimate (\ref{eq:eff_max}).

\subsubsection{Potential}

As an example, we consider the same potential as the one used in Ref. \cite{GLM}, even if the kinetic terms are now non canonical. In terms of the two scalar fields $\phi$ and $\chi$, the potential is given by 
\begin{eqnarray}
V(\phi,\chi) &=&
\frac12 M^2 \cos^2 \left(\frac{\Delta \theta}{2}\right)
\left[\chi - (\phi-\phi_0) \tan \Xi
\right]^2+
\frac12 m_\phi ^2 \phi^2\,,
\label{func_potential}
\end{eqnarray}
where the function $\Xi (\phi)$ is defined by
\begin{eqnarray}
\Xi (\phi) =  \frac{\Delta \theta}{\pi} \arctan [s (\phi - \phi_0)]\,.
\end{eqnarray}
The quantity $s$ is a constant parameter that controls the sharpness of the turn and is related to $\mu$ by
	\begin{align}
		\mu \simeq \sqrt{2}\, \dot{\phi}\, s \simeq \frac{2}{\sqrt{3}}c_s m_{\phi}s\,.
	\end{align}
The mass parameters $m_{\phi}$ and $M$ are chosen so that the condition
 $\sqrt{c_s}m_{\phi} \ll H \ll \sqrt{c_s}M$ is satisfied.

The potential (\ref{func_potential}) has been constructed so that the heavy direction, which  corresponds approximately to
\begin{eqnarray}
\phi_h \simeq \chi - (\phi - \phi_0)\tan \Xi\,,
\label{eq_valley}
\end{eqnarray}
changes its direction in  field space around $\phi=\phi_0,\;\;\chi=0$.

\subsubsection{Parameters}

Our model contains six parameters. The turn itself is characterized by three parameters,  $\{ \Delta \theta, s, \phi_0 \}$. The potential also depends on two mass parameters, $m_\phi$ and $M$, while the kinetic term depends on the parameter $f$. 
The Hubble scale $H$ can be fixed by requiring that the observed power spectrum, ${\cal P}_{\zeta0} \simeq 10^{-9}$ is approximately reproduced, for $\phi \simeq \phi_0$, in the limit $\Delta \theta \to 0$. This gives
	\begin{align}
		H &\simeq 2\pi M_P\sqrt{2c_s\, \epsilon \, {\cal P}_{\zeta0}} ,
	\end{align}
where we have explicitly written the Planck scale $M_P$. 
Moreover, approximating $V \simeq m_{\phi}^2 \phi^2/2$, 
we find from $\epsilon \simeq c_s \epsilon_V$,
	\begin{align}
		\phi_0 \simeq -M_P\sqrt{\frac{2c_s}{\epsilon}} ,
	\end{align}
and from Eq. (\ref{cs_approximation}),
	\begin{align}
		f \simeq \frac{c_s}{2\epsilon M_   P^2H^2}\left(\frac{1}{c_s^2}-1\right) .
	\end{align}
Finally, using the Friedmann equation (\ref{eom_bg}) within the slow-roll approximation, 
the light mass $m_{\phi}$ can be determined as
	\begin{align}
		m_{\phi} \simeq H\sqrt{\frac{3\epsilon}{c_s}} .
	\end{align}
To fix the parameters, we also assume the value of the slow-variation parameter $\epsilon (\simeq c_s\epsilon_V)$ to be $0.01$. Given these constraints between the various parameters, it is more convenient to parametrize our  system by the four parameters $\{\Delta \theta, s, M, c_s\}$.

\subsubsection{Efficiency: numerical result}
\begin{figure}[tbh]
		\begin{minipage}{.48\linewidth}
		\includegraphics[width=.98\linewidth]{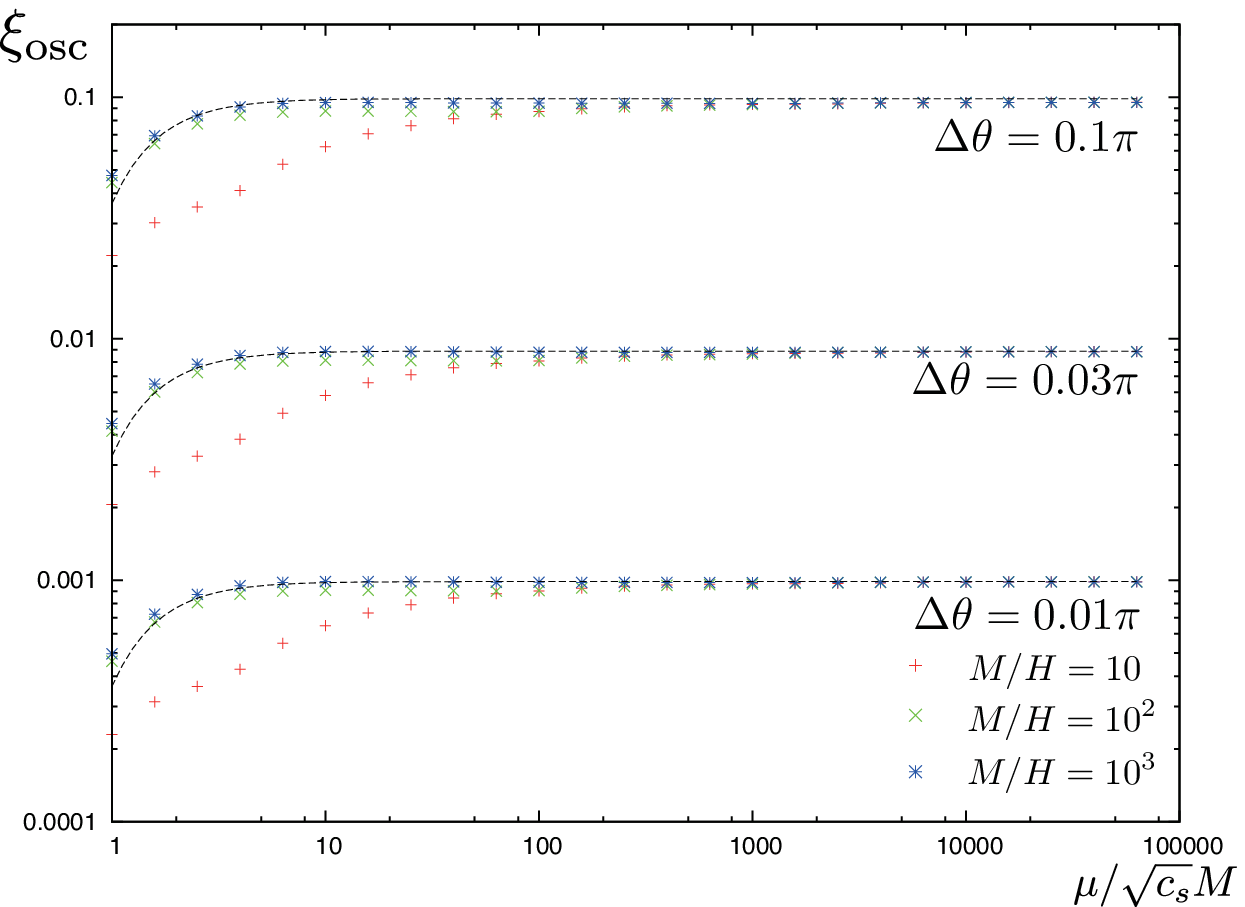}
		\end{minipage}
		\begin{minipage}{.48\linewidth}
		\includegraphics[width=.98\linewidth]{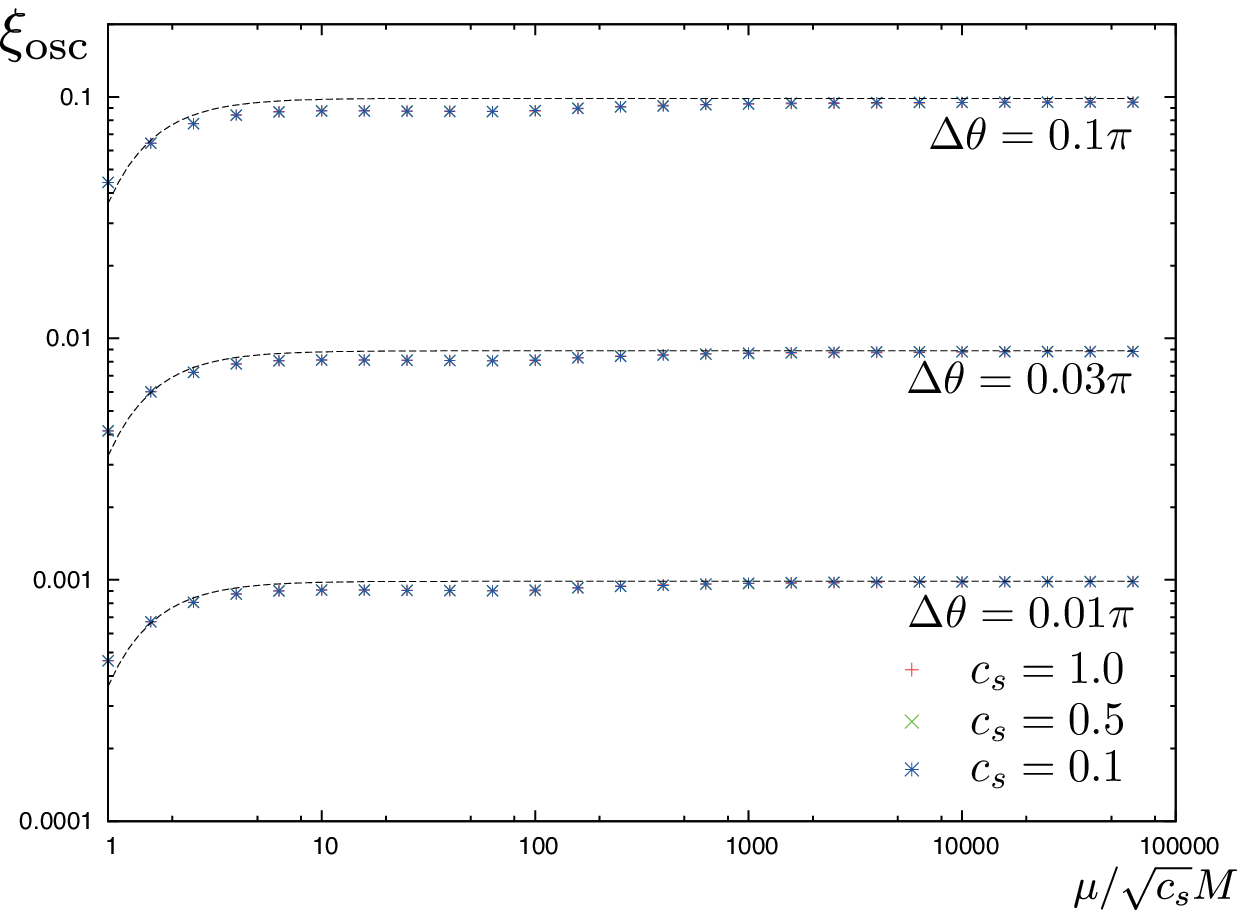}
		\end{minipage}
		\caption{Efficiency of the excitation of the heavy scalar field with respect  to the sharpness of the turn $\mu$ for various values of $M$ (left panel) and $c_s$
 (right panel). In each plot, the other parameter is fixed as $c_s=0.5$ and $M/H=10^2$, respectively. The points represent numerically estimated values of the efficiency, while the dotted lines correspond to the  analytical estimate (\ref{E_r_est}). 
Here, we have normalized the sharpness $\mu$ by $\sqrt{c_s}M$ since the condition for the sharp turn (\ref{eq:sharp_turn}) approximately corresponds to $\mu \gg \sqrt{c_s}M$ in the present setup.}
		\label{fig:efficiency}
	\end{figure}

 We have evaluated  numerically the efficiency parameter  (\ref{E_r_def}) for various values of the parameters $\{\Delta \theta, s, M, c_s\}$ by  solving the equations of motion (\ref{eqs_fields_simp}).
 In Fig.~\ref{fig:efficiency}, we have plotted our numerical estimate of the efficiency as a function of $\mu/\sqrt{c_s} M$, 
 for various values of the angle $\Delta \theta$, of the ratio $M/H$ and of the sound speed $c_s$.
 Comparing these results  with the analytical expression (\ref{E_r_est}), one finds that the latter provides a very good approximation provided the heavy mass is sufficiently large or the turn sufficiently sharp, i.e.
 $\mu \gg \sqrt{c_s}M$. In particular, one can check that the sharp-turn limit corresponds to $\Delta \theta^2$ and is indeed independent of the values of $M$ and $c_s$. 

The deviation from the analytical result when $\mu$ is not so large, can be explained by the dependence 
of the efficiency  on the details of the evolution of the angle $\theta_p$ during the turn.  Indeed, one does not expect the Gaussian approximation (\ref{thetapdot}) to be accurate in this regime.
Moreover, the assumption that $\sqrt{c_s}M$ is constant is also not valid during the turn.

In conclusion, while the efficiency depends on how the angle $\theta_p$ evolves during the turn when the turn is not very sharp, we have confirmed that 
in the sharp-turn limit, the efficiency only depends on the global variation of the angle, $\Delta \theta$.

\section{The features in the primordial spectra}
\label{sec:Features}
In this section, we first discuss  the features induced by the mixing effect in the power spectrum, then the features induced by the resonance effect both in the power spectrum and the bispectrum. 
Finally, we compare these two types of features and show that there exist simple relations between them, which depend only on  parameters that are principle measurable.


\subsection{Features induced by the mixing}

Let us briefly summarize the analysis of \cite{GLM} devoted to  the features generated by the mixing
caused by a sudden turn  of the inflationary trajectory,  for models with canonical kinetic terms

First of all, it is important to stress that this analysis uses the mass basis,  which consists of  the  eigenvectors of the effective mass matrix for the fluctuations. This effective mass matrix contains the second derivatives of the potential but also terms that depend on the time derivatives of the scalar fields, and therefore the mass basis in general does not coincide with the potential basis defined in (\ref{V_mn_diag}).
In models with canonical kinetic terms, it turns out that  the additional terms are slow-roll suppressed, so that  the potential basis provides an excellent approximation of the mass basis. 
By contrast, in models with non canonical kinetic terms, the deviation between the mass and potential bases can become important, in particular when the sound speed is small. 
In this paper, we assume that the sound speed remains large enough so that these effects are not significant and can be neglected. A refined treatment, including the cases with  a small sound speed, is left for  future work.

Since we are interested by the final curvature power spectrum, which is ultimately observable, we concentrate on the late-time power spectrum of the light mode on super-Hubble scales,  noting that the light and adiabatic direction  coincide sufficiently long  after the turn. 
The light and heavy modes being initially statistically independent, 
the final power spectrum of the light mode  can be expressed as  the sum
of two contributions,
\begin{eqnarray}
{\mathcal P}_{l,{\rm mix}}(k)={\mathcal P}_{l} ^{(l)}(k)
+{\mathcal P}_l ^{(h)}(k)\,.
\end{eqnarray}
The first contribution, ${\mathcal P}_{l} ^{(l)}$, which we refer to as the  light contribution,  is obtained with  initial conditions where
 the light mode is in its Bunch-Davies vacuum state and the heavy mode is zero.
The second contribution, ${\mathcal P}_{l} ^{(h)}$, or heavy contribution, is obtained with 
 initial conditions where the heavy mode is in its Bunch-Davies vacuum state
and the light mode is zero.

In a perturbative treatment, where $\Delta\theta$ is the small parameter,  the  light contribution can be formally written as 
\begin{eqnarray}
{\mathcal P}_l ^{(l)}(k)=
\left(1+\mathcal{F}_l + \mathcal{F}_{lh}\right){\mathcal P}_{l0} +\dots
\end{eqnarray}
The zeroth order term, ${\mathcal P}_{l0}$,   is 
the power spectrum without any turn (i.e. $\Delta \theta=0$).
On top of it, one finds two  corrections of order $\Delta\theta^2$. 
The first one, denoted $\mathcal{F}_l$, arises from the self-coupling of the light mode, proportional to  $(\Delta \theta)^2$. The second correction, denoted $\mathcal{F}_{lh}$, comes from  the coupling between the light and heavy mode, proportional to $\Delta \theta$: even though there is no heavy mode initially, this coupling generates 
a heavy mode with  amplitude $\sim \Delta \theta$, which in turn induces  a  correction of order
$(\Delta \theta)^2$ in the light mode. 

A similar analysis can be conducted for the heavy contribution.  With a non zero heavy mode initially, the coupling between the light mode and heavy mode  generates 
a light mode of amplitude $\sim \Delta \theta$.  Since  the heavy contribution vanishes in the absence of  mixing, this will induce a correction of  order $(\Delta \theta)^2$ in the power spectrum, which we denote as $\mathcal{F}_h$, so that 
\begin{eqnarray}
{\mathcal P}_l ^{(h)}(k)=\mathcal{F}_h {\mathcal P}_{l0}.
\end{eqnarray}

Putting everything together,  the features in the power spectrum generated
by the mixing are given by
\begin{eqnarray}
\label{eq:feature_mix_form}
\frac{\Delta {\mathcal P}_{l,{\rm mix}}}{{\mathcal P}_{l0}}
\equiv \frac{{\mathcal P}_{l,{\rm mix}}}{{\mathcal P}_{l0}}-1
=\mathcal{F}_l +\mathcal{F}_h+ \mathcal{F}_{lh},
\label{exp_feature_mix}
\end{eqnarray}
all terms being of order $\sim (\Delta \theta)^2$.
 The amplitude of these corrections  peaks around the scale
that crosses the Hubble scale at the time of the turn $t_*$, i.e.
$k_{\rm mix} \equiv a_* H$ with $a_* \equiv a(t_*)$, and then decreases 
as $k$ increases. 
In fact, as pointed out in \cite{Konieczka:2014zja}, the sum of the three terms
in Eq.~(\ref{exp_feature_mix}) vanish in the limit  $k \to \infty$.

In the case of a sharp turn,  the term  $\mathcal{F}_h$ dominates on scales around
 $k_{\rm mix}$, as $\mathcal{F}_l$ and $\mathcal{F}_{lh}$ tend  to cancel each other on scales
$H < k/a_* < m_h$,  as shown in \cite{Noumi:2013cfa}. We thus need to consider only this term, whose explicit form is given by
\begin{align}\label{eq:feature_mix_int}
		\mathcal{F}_h =  \lim_{k|\tau| \ll 1}\left|\int^{\tau}{\rm d}\tau' ~ G_l(\tau, \tau')\{\theta_p'' u_{h0}(\tau') + 2\theta_p' u_{h0}'(\tau') \}\right|^2\,.
	\end{align} 
In the above expression $\tau$ denotes the conformal time and 
$G_l(\tau',\tau)$ is the Green's function for the light mode,
\begin{align}
		G_l(\tau',\tau) = 2\Im\{ u_{l0}(\tau')u_{l0}^{\ast}(\tau) \},
	\end{align}
	where $u_{m0}$ ($m=l,h$) denote the Bunch-Davies solutions 
of the canonically normalized fluctuations $u_m \equiv a \delta \phi_m$, explicitly 
given by  
\begin{eqnarray}
u_{l0}(\tau) &=& \frac{e^{-ik\tau}}{\sqrt{2k}}\left( 1- \frac{i}{k\tau} \right),\\
u_{h0} (\tau) &=& \frac{\sqrt{\pi}}{2}e^{-\frac{\pi}{2}\nu+i\frac{\pi}{4}}\sqrt{-\tau}H^{(1)}_{i\nu}(-k\tau), \quad \nu \simeq \frac{m_h}{H},
\end{eqnarray}
where $H^{(1)}$  is the Hankel function of the first kind. 

It can then be shown that the maximum amplitude of the features, around the scale $k_{\rm mix}$, is given by
	\begin{align}
		\frac{\Delta {\cal P}_{\zeta, {\rm mix}}}{{\cal P}_{\zeta0}} 
&\sim (\Delta \theta)^2e^{-\frac{m_h^2}{\mu^2}}\left(\frac{m_h}{H}\right) \\
		&\sim \xi_{\rm osc}\left(\frac{m_h}{H}\right).
		\label{eq:feature_mix}
	\end{align}

In  DBI inflation,  the structure of the evolution equations
for the perturbations is modified with respect to models with canonical kinetic terms
and the features in the power spectrum are thus   different in principle.
However, when the sound speed is not so different from unity
and when the derivative interactions remain small enough, it is natural to expect that the corrections to the features generated by the derivative interactions are  not significant and that 
Eq.~(\ref{eq:feature_mix}) remains quantitatively adequate to estimate  the maximum amplitude of the features.
In our specific model, we have checked numerically that this is indeed true for $c_s \geq 0.7$ 
with fixed $m_h$ and $\mu$,  although we observe a  shift of 
$k_{\rm mix}$ from $a_\ast H$ to $a_\ast H/c_s$ which can be easily explained 
by the modification of the sound horizon. On the other hand,
for very small $c_s$, we observe  specific effects due to derivative interactions
in the features of the power spectrum, such as  an enhancement of  $\mathcal{F}_h$ by a factor $\sim 1/c_s$.
As  mentioned previously,  we will assume here that the sound speed is 
close enough to  unity
that the maximum amplitude of the features is well approximated by 
Eq.~(\ref{eq:feature_mix}) and leave the detailed analysis of the perturbations
for  small sound speeds for a future work.

\subsection{Features induced by the resonance}
Since the DBI Lagrangian (\ref{multifieldaction}) contains derivative interactions, 
the inflaton fluctuations can be efficiently amplified through the resonance with the oscillations in the background trajectory \cite{Saito:2012pd}.
When the sound speed is not so small ($f\tilde{X} \ll 1$), the square root in the Lagrangian (\ref{multifieldaction}) can be expanded as
	\begin{align}\label{eq:DBI_expand}
		P(X^{IJ},\phi^I) &= X_{11}+X_{22} + \frac{f}{2}\left[ (X_{11} - X_{22})^2 + 4X_{12}^2 \right]  \nm \\ 
		& \qquad\qquad + \frac{f^2}{2}(X_{11}+X_{22})\left[ (X_{11} - X_{22})^2 + 4X_{12}^2 \right] + \cdots\,.
	\end{align}
	Expressing the above expression in the mass basis $(\phi_l, \phi_h)$, one finds that 
the action contains  derivative interactions between the light and heavy modes of the form
	\begin{align}\label{eq:kd}
		\frac{\lambda_{d1}}{4\Lambda_d^4}(\pd \phi_l)^2(\pd \phi_h)^2, \qquad \frac{\lambda_{d2}}{4\Lambda_d^4}(\pd \phi_l \cdot \pd \phi_h)^2, 
	\end{align}
with $\Lambda_d=f^{-1/4}$, $\lambda_{d1}=-1$, and $\lambda_{d2}=2$, as well as the self-interaction term 
	\begin{align}\label{eq:ks}
		\frac{\lambda_s}{4\Lambda_d^4}(\pd \phi_l)^4,
	\end{align}
with $\lambda_s=1/2$ and higher-order terms, 
	\begin{align}\label{eq:kh}
		\frac{\lambda_{h1}}{8\Lambda_d^8}(\pd \phi_l)^4(\pd \phi_h)^2, \qquad \frac{\lambda_{h2}}{8\Lambda_d^8}(\pd \phi_l)^2(\pd \phi_l \cdot \pd \phi_h)^2 ,
	\end{align}
with $\lambda_{h1}=-1$ and $\lambda_{h2}=2$, which have been considered 
in Ref.~\cite{Saito:2012pd}. 

The feature in the power spectrum induced by the resonance has the largest amplitude around the scale that crosses the mass scale at the time of the turn, $k_{\rm res} \equiv a_{\ast}m_h$, with
	\begin{align}\label{eq:feature_res_pow}
		\frac{\Delta {\cal P}_{\zeta, {\rm res}}}{{\cal P}_{\zeta0}} \sim -\frac{1}{4}( \lambda_{d1} + 2\lambda_{d2} )q_d\sqrt{\frac{m_h}{H}}; \quad q_d \equiv f\dot{\phi}_{h, \ast}^2.
	\end{align}
In the case of the DBI action, 
the above  expression accidentally vanishes.
However, 
if we take into account the reduction of the sound speed due to the self-interaction term (\ref{eq:ks}), 
it can be shown that the correction has a non-vanishing contribution (see Appendix \ref{sec:app}):
	\begin{align}\label{eq:feature_res_pow_next}
		\frac{\Delta {\cal P}_{\zeta, {\rm res}}}{{\cal P}_{\zeta0}} \sim (1-c_s^2)q_d\sqrt{\frac{m_h}{H}} .
	\end{align}
The parameter $q_d$ represents the strength of the derivative interactions and can be written in terms of the sound speed (\ref{eq:c_s}) as,
	\begin{align}
		q_d &= f\dot{\sigma}^2 \xi_{\rm osc} \\
		&= \left(\frac{1}{c_s^2} - 1\right)\xi_{\rm osc} \simeq (1-c_s^2)\xi_{\rm osc},
	\end{align}
where we have used $c_s \simeq 1$ in the last equality. 
The expression (\ref{eq:feature_res_pow_next}) was obtained by using the perturbation for the derivative interactions.
The perturbation is only valid when $\Delta {\cal P}_{\zeta, {\rm res}}/{\cal P}_{\zeta0} < 1$, or
	\begin{align}\label{eq:perturbativity_res}
		c_s^2 > \frac{1}{1+(\xi_{\rm osc}\sqrt{\frac{m_h}{H}})^{-\frac{1}{2}}}.
	\end{align}
Here, we assume that $c_s$ satisfies the condition (\ref{eq:perturbativity_res}) 
and that we can use the result (\ref{eq:feature_res_pow}) for the correction in the power spectrum due to the resonance effect.
Since $\Delta {\cal P}_{\zeta, {\rm mix}}/{\cal P}_{\zeta 0}<1$ indicates $\xi_{\rm osc}\sqrt{m_h/H} < \sqrt{H/m_h} \ll 1$, 
this condition is satisfied unless the features by the mixing is prominently large. 

Features are also induced in the bispectrum. 
Introducing the dimensionless quantity ${\cal B}_{\zeta}$ defined by
	\begin{align}\label{eq:dlB}
		\langle \zeta_{{\bf k}_1}(t)\zeta_{{\bf k}_2}(t)\zeta_{{\bf k}_3}(t) \rangle \equiv (2\pi)^7\delta^3({\bf k_1+k_2+k_3}){\cal P}_{\zeta 0}^2\frac{{\cal B}_{\zeta}}{k_1^2k_2^2k_3^2},
	\end{align}
the feature reaches its  largest amplitude around the scales $K \equiv k_1+k_2+k_3 \simeq 2k_{\rm res}$, with
	\begin{align}\label{eq:feature_res_bi}
		\Delta {\cal B}_{\zeta,{\rm res}} \sim \max\left\{ \epsilon q_d \left(\frac{m_h}{H}\right)^{\frac{3}{2}}, (1-c_s^2)q_d \left(\frac{m_h}{H}\right)^{\frac{5}{2}} \right\}.
	\end{align}
Here, 
``$\max$" indicates that one should take the larger term in the curly brackets because the correction to the bispectrum has contributions from both the second-order and fourth-order terms in the Lagrangian (\ref{eq:DBI_expand}).
In the case of the bispectrum, 
there is no accidental cancellation similar to that found  in the power spectrum.

\subsection{Relation between the features}\label{ss:relation}
Given the  explicit expressions for the  features  in terms of the main parameters, it is now easy to find relations between them. The comparison of  Eqs. (\ref{eq:feature_mix}), (\ref{eq:feature_res_pow_next}), and (\ref{eq:feature_res_bi}) yields
	\begin{align}
		\frac{\Delta {\cal P}_{\zeta, {\rm res}}}{{\cal P}_{\zeta0}} &\sim (1-c_s^2)^2 \left(\frac{m_h}{H}\right)^{-\frac{1}{2}}\frac{\Delta {\cal P}_{\zeta,{\rm mix}}}{{\cal P}_{\zeta0}}, \label{eq:rel_pow}\\
		\Delta {\cal B}_{\zeta,{\rm res}} &\sim \max\left\{ (1-c_s^2)\epsilon\left(\frac{m_h}{H}\right)^{\frac{1}{2}}, (1-c_s^2)^2\left(\frac{m_h}{H}\right)^{\frac{3}{2}} \right\}\frac{\Delta {\cal P}_{\zeta,{\rm mix}}}{{\cal P}_{\zeta0}}.\label{eq:rel_bi}
	\end{align}
 These expressions show that the resonance cannot induce a large feature in the power spectrum, but only in the bispectrum. 
 The relation (\ref{eq:rel_pow}) is only valid if the sound speed is not 
 too small since we have assumed $c_s \geq 0.7$  for the mixing, as well as  the condition (\ref{eq:perturbativity_res}) for the resonance.
The latter condition is automatically satisfied if the resonance feature has a reasonable amplitude, $\Delta {\cal P}_{\zeta, {\rm res}}/{\cal P}_{\zeta0} < 1$. 
 Thus, the condition $c_s \ge 0.7$ is sufficient to ensure the validity of the relation (\ref{eq:rel_pow}) as far as we consider reasonable modulations in the power spectrum. 
 This condition is not so restrictive to get a large modulation in the bispectrum since the  factor $m_h/H$ can be sufficiently large to compensate the suppression by the factor $(1-c_s^2) \le 0.5$ for $c_s \ge 0.7$.

Note that the amplitude (\ref{eq:rel_bi}) is written only in terms of observable parameters. 
The sound speed and the slow-variation parameter can be deduced in principle from the observations of the tensor modes and equilateral non-Gaussianity: $r = 16c_s\epsilon$, $n_t=-2\epsilon$, and $f_{\rm NL} \sim -(35/108)(c_s^{-2}-1)$.
Moreover, $m_h/H$ can be determined by the relative position of the two features, according to
	\begin{align}
		\frac{m_h}{H} = \frac{k_{\rm res}}{k_{\rm mix}}.
	\end{align}
Therefore, given a mixing feature observed in the power spectrum, one could look for a corresponding  resonance feature in the bispectrum. The observation of such combination of features in the data
would provide a strong evidence for the presence of a heavy mode, excited during inflation.

\section{Summary}
In this paper, we have analysed a sharp bending of the trajectory in two-field DBI inflation, 
where two types of features in the primordial spectra are expected to be induced, due to mixing and resonance effects, respectively. Indeed, in addition to  features induced by a large mixing between the light and heavy modes, derivative interactions, which are contained in the non-canonical kinetic term of the DBI action, can lead to an efficient enhancement of the fluctuations through a resonance with the oscillations of the heavy field excited by a sharp turn. 

To see the relation between these two features, 
we have investigated how the amplitude of the oscillations is determined by the details of the turn, taking into account the effect of the derivative interactions, i.e. the reduction of the sound speed.
Introducing the efficiency parameter, defined as the ratio between the maximum kinetic energy of the heavy field and the total kinetic energy before the turn, 
we found that it is determined only by the global variation of the light direction angle in the sharp-turn limit.

With the assumption that the sound speed remains rather large ($c_s \ge 0.7$), we find   that the maximum amplitude of the both mixing and resonance features does not depend on the information of the turn other than the efficiency parameter. 
Remarkably, it is then possible to derive consistency relations between the two types of features.
As consequences, 
it was shown that the resonance cannot induce a large feature in the power spectrum, 
but only in the bispectrum due to a larger amplification factor of the heavy mass.
The resultant resonance features are suppressed by a factor of $1-c_s^2$, 
which represents the strength of the derivative interactions. 
However, 
even when the kinetic term can be approximated to be canonical, $c_s \simeq 1$, 
the amplification factor is large enough to realize a non-negligible feature in the bispectrum 
when the heavy mass is much larger than the Hubble scale.

If the combination of both mixing and resonance features were detected and shown to satisfy the consistency relation, it would give a compelling evidence for the existence of a heavy field during inflation. 
Moreover, the relative position of the features would  give directly the value of the heavy mass. It would thus be interesting to look for such features in the data, following techniques that have been developed 
in e.g. Refs. \cite{Flauger:2009ab,Meerburg:2011gd,Chen:2012ja,Jackson:2013mka,Ade:2013uln,Meerburg:2013cla,Meerburg:2013dla,Clesse:2014pna,Chen:2014joa,Munchmeyer:2014nqa}. 

In this paper,  we have not discussed the cases with a small sound speed $c_s < 0.7$. 
None of the current data point to a small sound speed, but there is  still room for it, since the current constraints on  equilateral non-Gaussianity give the lower bound  $c_s > 0.07$ at 95\% CL \cite{Ade:2013ydc} (assuming a single light field).
As already mentioned, the results of the present work do not apply for a small sound speed and further investigations are required. 
We will study these cases in a future work.

\acknowledgements
D.L. was partly supported by ANR (Agence Nationale de la Recherche) grant ``STR-COSMO" ANR-09-BLAN-0157-01. S.M. is grateful to the APC for their
hospitality when this work was almost done. 
R.S. is supported by Grant-in-Aid for JSPS postdoctoral fellowships for research abroad.

\appendix

\section{Resonance features in the power spectrum for the DBI action}\label{sec:app}
In the case of DBI, it turns out that 
the correction  to the power spectrum due to the resonance, 
which has been obtained by a perturbative expansion with respect to the derivative interactions, 
accidentally vanishes. 
In this appendix, we show that the correction is nevertheless non-vanishing  
if one takes into account the reduction of the sound speed  due to the self-interaction term (\ref{eq:ks}) included in the DBI action.

To see the general structure of the correction, 
we first calculate it without fixing the values of $\lambda_{d1}$, $\lambda_{d2}$, and $\lambda_s$, 
keeping in mind that 
	\begin{align}\label{eq:DBIlambda}
		\lambda_{d1}=-1, \quad \lambda_{d2}=2, \quad \lambda_s=\frac{1}{2},
	\end{align}
for the DBI action.
For brevity, 
we introduce the quantity that controls the amplitude of the self-interaction term,
	\begin{align}
		q_s \equiv f\dot{\phi}_l^2.
	\end{align}
Comparing the definition of $q_d$, 
which controls the amplitude of the interaction between the light and heavy modes, 
we find
	\begin{align}
		\frac{q_d}{q_s} = \frac{\xi_{\rm osc}}{1-\xi_{\rm osc}}.
	\end{align}
Hence, though $q_s$ is slow-roll suppressed, it has a value larger than $q_d$. 
It contributes to the non-oscillatory part of the sound speed as
	\begin{align}
		\bar{c}_s^2 = \frac{1+\lambda_s q_s}{1+3\lambda_s q_s}.
	\end{align}
When the resonance is not relevant, 
it gives the leading contribution to the sound speed.

The quadratic Hamiltonian in this system is given by 
\cite{arXiv:0806.0336,Arroja:2008yy},
	\begin{align}
		H^{(2)} = \int \mr{d}^3x~ \frac{1}{2}\left[ \dot{v}^2 + \left(\frac{c_s^2\nabla^2}{a^2} - \frac{\ddot{z}_{\phi}}{z_{\phi}} \right)v^2 \right] . \label{eq:2hamiltonian}
	\end{align}
where 
	\begin{align}
		z_{\phi}^2 &\equiv a^3(2P_{1K,L1}X^{KL}+P_{11}), \\
		c_s^2 &\equiv \frac{P_{11}}{2P_{1K,L1}X^{KL}+P_{11}} , 
	\end{align}
and $v\equiv z_{\phi}\delta \phi_l$.
Extracting the leading-order oscillatory components, 
we find
	\begin{align}
		z_{\phi}^2 &= a^3\left[1 + 3\lambda_s q_s + (\lambda_{d1}+\lambda_{d2})q_d\sin^2(m_h t) \right], \label{eq:zphi_osc}\\
		c_s^2 &= \bar{c}_s^2\left[1 + \frac{1}{1+3\lambda_{s}q_s}\left(\frac{\lambda_{d1}}{\bar{c}_s^2} - \lambda_{d1}-\lambda_{d2} \right)q_d\sin^2(m_h t)  \right] + {\cal O}(q_d^2). \label{eq:cs_osc}
	\end{align}
Here, 
we did not perform an expansion with respect to $q_s$ unlike the treatment in Ref. \cite{Saito:2012pd} to see the effect of the reduction in the sound speed due to the self-interaction, $\bar{c}_s^2 < 1$.

Using the method of steepest descent, 
it can be shown that the peak amplitude of the correction to the power spectrum is given by
	\begin{align}
		\frac{\Delta {\cal P}_{\zeta, {\rm res}}}{{\cal P}_{\zeta 0}} \sim \left.\left(\frac{c_s^2}{\bar{c}_s^2} - \frac{\ddot{z}_{\phi}}{z_{\phi}}\right)\right|_{\rm osc}\sqrt{\frac{m_h}{H}} ,
	\end{align}
where ``osc" in the subscript indicates that one should pick up the oscillatory components. 
From Eqs. (\ref{eq:zphi_osc}) and (\ref{eq:cs_osc}), 
it can be estimated as
	\begin{align}
		\left.\left(\frac{c_s^2}{\bar{c}_s^2} - \frac{\ddot{z}_{\phi}}{z_{\phi}}\right)\right|_{\rm osc} \simeq -\frac{q_d}{2(1+3\lambda_s q_s)}\left( \frac{\lambda_{d1}}{\bar{c}_s^2} +\lambda_{d1} + \lambda_{d2} \right) ,
	\end{align}
where we have discarded terms suppressed by $H/m_h$.
Then, substituting the DBI values given in  (\ref{eq:DBIlambda}), 
	\begin{align}
		\left.\left(\frac{c_s^2}{\bar{c}_s^2} - \frac{\ddot{z}_{\phi}}{z_{\phi}}\right)\right|_{\rm osc} &\simeq \frac{q_d}{2(1+3\lambda_s q_s)}\left( \frac{1}{\bar{c}_s^2} - 1 \right) ,\\
		&\simeq \frac{q_d}{2}(1-\bar{c}_s^2),
	\end{align}
where we have used $q_s \ll 1$ and then $\bar{c}_s^2 \simeq 1$ in the last equality.
Therefore, although
the resonant feature vanishes at the leading order of the derivative interactions in the DBI case, 
 it is no longer the case  if one takes into account the reduction of the speed sound due to the self-interaction term.

\end{document}